\documentclass[pra,showpacs,superscriptaddress,twocolumn]{revtex4}
\usepackage{bm,graphicx,amsmath}
\usepackage{amsfonts,amssymb}

\begin{document}

\title{Quantum interference structures in trapped ion dynamics beyond the Lamb-Dicke and rotating wave approximations}

\author{Dong Wang}
\author{Tony Hansson}
\author{\AA sa Larson}
\affiliation{Department of Physics, Stockholm University, AlbaNova University Center, SE-106 91 Stockholm, Sweden}
\author{Hans Karlsson}
\affiliation{Department of Physical and Analytical Chemistry, Quantum Chemistry, Uppsala University, Box 518, SE-751 20 Uppsala, Sweden}
\author{Jonas Larson}
\affiliation{ICFO-Institut de Ci\`{e}ncies Fot\`{o}niques, E-08860 Castelldefels, Barcelona, Spain}
\date{\today}

\begin{abstract}
We apply wave packet methods to study an ion-trap
system in the strong excitation regime imposing neither the rotating
wave nor the Lamb-Dicke approximations. By this approach we show the
existence of states with restricted phase space evolution, as a
genuine consequence of quantum interference between wave packet
fractions. A particular instance of such a state oscillates between
maximal entanglement and pure disentanglement between the constitute
subsystems, where the characteristic crossover time is very rapid. Over longer time periods the dynamics of these states
exhibits collapse-revival patterns with well resolved fractional
revivals in autocorrelation, inversion and entanglement.
\end{abstract}

 \maketitle

\section{Introduction}
Progress in research fields as laser cooling, production and
controlling sub femtosecond laser pulses, manufacturing of solid
state devices, and so forth, render experiments of novel quantum
characters. A key model in quantum optics is the harmonically
trapped ion pumped by external lasers \cite{trapreview}, which
successfully has been used to study pure quantum phenomena in
experiments \cite{expcat1,traprev,trapent,expcat2}. Both the
internal electronic structure of the ion and its center-of-mass
motion are treated quantum mechanically and it is possible to tune
experimental parameters such that a single or few electronic
transitions in the ion can be isolated and coherently coupled to the
motion. The spatial profile of the driving laser, either a standing
wave (SW) or a traveling wave (TW), effectively reshape the trapping
potentials. In particular, a unitary transformation of the regular
one-dimensional ion Hamiltonian with TW pumping may result in two
equally shifted and displaced coupled harmonic oscillators
\cite{hector1,hector2}, whereas SW fields in general generate
coupled potentials with a more complex shape. In addition, a fairly
new proposal, verified by experiments, is to place the trap inside a
high-$Q$ micro cavity such that it is the cavity field driving the
trapped ion \cite{expcav,theocav}. Even Bose-Einstein condensates
have been successfully trapped and coherently coupled to a single
cavity mode \cite{bec-cav}.

In most theoretical work, pertaining to both of the SW and TW cases,
approximations are imposed in order to obtain analytical or
semi-analytical results valid in different parameter regimes:
\begin{enumerate}
 \item \emph{The Lamb-Dicke} (LD) \emph{regime}. In this regime the
 wavelength of the classical laser field is long compared to the
 extent of the confining harmonic trap and an expansion in the
 small \emph{Lamb-Dicke parameter} $\eta$ of the mode profile function
 is made \cite{ldprep,ldprep2,ldprep3,ldrev,ldrev2,ldmeas}.

\item \emph{The rotating wave approximation} (RWA) \emph{regime}.
Here, a basis is defined in which the involved time scales differ
considerably and, once in the interaction picture, fast oscillating
terms are neglected \cite{rwasw,rwa2,rwaprep1,rwaprep2,rwaprep3}.
This assumes that the particular system can be separated into a
``slow'' and a ``fast'' part, which puts constraints on the
constituent frequencies; trap frequency $\omega$, ion transition
frequency $\Omega$, laser frequency $\omega_L$ and Rabi frequency
$\lambda$ (laser-ion coupling amplitude).

\item \emph{Strong and weak excitation regimes}. These are the
limiting situations of either a very large or small dimensionless
parameter $\lambda/\omega$ \cite{low,lowprep,strong} corresponding
to, respectively, strong or weak pumping of the ion. It should be noted that this definition assumes a moderate number of phonon excitations, othervice the effective Rabi frequency may become relatively large even in the weak excitation regime. This in fact, is the case of the current work.
\end{enumerate}
Typically, the parameter regimes for which the above approximations
are justified overlap and, in particular, there are regions in which
none of them is valid. In many studies, one of the approximations is
applied and one considers dynamics ``beyond'' the others. This,
however, can be misleading since one is often in a parameter regime
where also the non-imposed approximations could in principle have
been implemented. Recently, Liu et al. \cite{ground} showed that the
ground state of a TW pumped trapped ion may be considerably lowered
when both the RWA and the LD approximation are invalid. In the
current work we use a fully numerical method and therefore no
approximations are imposed. We discuss, nevertheless, the various
parameter regimes as well as introduce an additional approximation,
namely, the \emph{adiabatic approximation}, to provide further
insight into the quantum dynamics. It is commonly thought that the
dynamics, in contrast to the simple cases of the validity regimes
discussed above, becomes irregular beyond these approximations. We
will here show that this is indeed not always true.

Theoretical and experimental research on trapped ions has been
concerned with both SW and TW driving, with the main focus on the
latter. Examples include state preparation of non-classical
vibrational states such as Schr\"odinger cat states or Fock states
for both TW \cite{ldprep,lowprep,rwaprep1,rwaprep3,strong} and SW
\cite{ldprep2,ldprep3,rwaprep2,swprep2} pumping, collapse-revivals
in the TW \cite{traprev,ldrev2} and SW \cite{ldrev} cases, state
measurement \cite{ldmeas} and quantum information processing
\cite{trapent,qi}.

The above-mentioned theoretical works almost exclusively employed
the formalism of vibrational creation and annihilation operators for
the harmonic trap. We, on the other hand, will reformulate the model
in terms of ionic center-of-mass position and momentum, casting the
problem into one of two coupled harmonic oscillators. In particular,
we will study the coupled dynamics of an ion wave packet evolving on
the two potentials in the case of SW pumping. For a one-dimensional
system, as the one we consider, it resembles an idealized diatomic
molecule for which only two coupled bound electronic states are
taken into account. A similar analogy, but between a diatomic
molecule and a cavity QED model, was pointed out in
\cite{oldwine,jonaswp}, works that also made use of the wave packet
method. As the standard viewpoints and methods applied in quantum
optics and molecular physics are rather different, we discuss them
and their relation in some depth with the aim to facilitate
branching out and combining the two research fields.

\begin{figure}[!ht]
\begin{center}
\includegraphics[width=7cm]{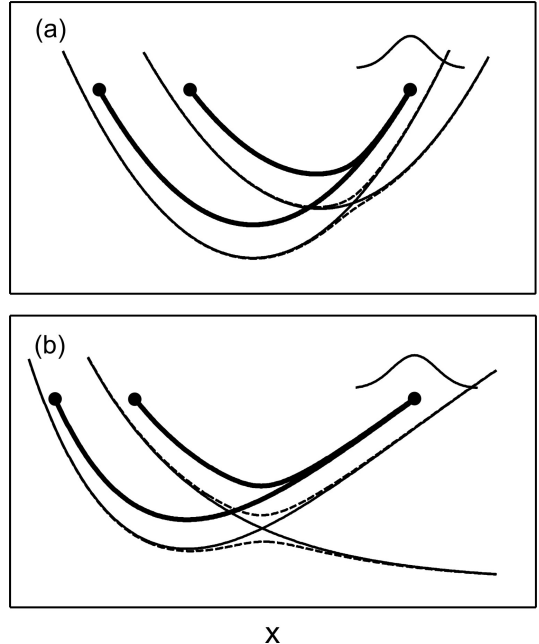}
\vspace{0.5cm} \caption{Schematic illustration of bistable wave
packet motion: the wave packet starts out in one diabatic state,
splits at the  crossing point and returns to the crossing where the
parts interfere to reproduce the initial wave packet after one
classical period of an individual potential well. The upper plot (a)
displays the example of two bound states, typical of our trapped ion
system, while the lower one (b) shows a bound state coupled to a
repulsive one, which is a system of relevance to molecular dynamics
(like (a)), where $x$ represents the internuclear distance.}
\label{fig1}
\end{center}
\end{figure}

Since the pioneering work in the seventies by Heller
\cite{heller75}, wave packet techniques have been used especially in
molecular and chemical physics research \cite{wpreview}. It is not
always possible to separate the dynamics in isolated electronic
states and the evolution on coupled electronic potentials has to be
considered; couplings could be of e.g. vibronic, spin-orbit, or
rotational nature or induced by an external laser field. In
Fig.~\ref{fig1}, we display two schematic examples of coupled
potentials, bound-bound in (a) and bound-repulsive in (b). Solid
potential curves are diabatic, while the corresponding adiabatic
ones are dashed (for a proper definition, see Sec. \ref{ssec2b}).
When the wave packet propagates on the coupled potentials,
interference effects will occur. In the bound-repulsive system, one
such effect manifests itself as unexpectedly long-lived vibrational
states \cite{dietz96}. This takes place when the interference
restricts the wave packet evolution in such a way that the repulsive
part of the system at the right side of the curve crossing is not
reached, indicated with the heavy solid lines in Fig.~\ref{fig1}.b.
We call this a \textit{bistable} motion
\cite{zhang,gado1,gado2,dong}. Similar interference effects can also
occur in a system consisting of two coupled bound states, in which
case the wave packet always returns to the same diabatic potential
after one oscillation (see Fig~\ref{fig1}.a). In \textit{astable}
motion (not shown), the wave packet starts in one diabatic state,
splits at the curve crossing and when it returns to the curve
crossing it recombines in such a way that it switches completely to
the other diabatic state. Consequently, astable motion cannot exist
in a system comprising coupled bound-repulsive states. We have
previously carried out a numerical wave packet study of the dynamics
of bistable trajectories \cite{dong} and found that the trajectories
can be exceedingly long-lived with sharp fractional and full
revivals.

Here we extend our former work and explore the dynamics of a system
describing a trapped ion driven by a standing wave. We point out
that also in such a system bistable motion exists and investigate
the short- and long-term evolution of the system under this
interference condition.

The outline of the article is as follows. We begin Sec. \ref{sec2}
by presenting the trapped ion model Hamiltonian, as it is generally
stated in quantum optics, and discuss parameter regimes and the
corresponding approximations -- rotating wave, Lamb-Dicke, weak and
strong excitation and adiabatic. It is shown how the Hamiltonian
relaxes to an effective one mimicking the acclaimed Jaynes-Cummings
(JC) model of quantum optics \cite{jc}. The relation between
frequently used bases are considered in Sec. \ref{ssec2b} for which
we define the corresponding potential curves. This enables us to
obtain a deeper intuition about the dynamics and physical properties
of the coupled system. The analogous molecular electronic states
with their potential energy curves and couplings are also discussed,
briefly. The following Sec. \ref{sec3} is dedicated to our numerical
results. First we define the concept of bistable motion and then
show how it enters into the trapped ion model. For short
timescales, we show in Sec. \ref{ssec3a} that this type of dynamics
oscillates between an entangled (can be made maximally entangled)
and a disentangled state. We analyze the nature of the wave packet
evolution around the potential curve crossing in terms of the ionic
inversion, the autocorrelation and the von Neumann entropy. For very
long times (Sec. \ref{ssec3b}) we observe distinct wave packet
collapse, fractional and full revival structures. Finally, Sec.
\ref{sec4} gives a short summary.

\section{The model system}\label{sec2}
This section introduces the model trapped-ion Hamiltonian. The
problem is first formulated in terms of operator algebra, commonly
used in the quantum optics community, and then recast in the
conjugate variable representation frequently applied in, for
instance, molecular physics. Within the latter representation, the
obtained Hamiltonian resembles that of an artificial diatomic
molecule with two coupled bound electronic states. There are several
parameters determining the particular form of the potential curves
and coupling, namely, effective ion-laser coupling $\lambda$ (Rabi
frequency), trap frequency $\omega$, laser frequency $\omega_L$, ion
transition frequency $\Omega$, LD parameter (laser wavelength)
$\eta$, and laser phase shift $\phi$. For a large LD parameter, the
curves exhibit several crossings, whereas for small $\eta$ one may
encounter only a few, one or even no crossings in the parameter
range of interest. We will here be interested exclusively in the
situation
 of a single level crossing of the two potential curves.

\subsection{Trapped ion in a standing wave}\label{ssec2a}
In the formalism of phonon creation $\hat{a}^\dagger$ and
annihilation $\hat{a}$ operators the Hamiltonian reads (we use
atomic units so that $\hbar=1$)
\begin{equation}\label{ham1}
H=\omega\hat{a}^\dagger\hat{a}+\frac{\Delta}{2}\hat{\sigma}_z+\lambda
\cos[\eta(\hat{a}^\dagger+\hat{a})+\phi]\hat{\sigma}_x,
\end{equation}
with the ion-laser detuning $\Delta=\Omega-\omega_L$ and the
$\hat{\sigma}$-operators act on the two ionic levels $|1\rangle$ and
$|2\rangle$, $\hat{\sigma}_z=|2\rangle\langle 2|-|1\rangle\langle
1|$ and $\hat{\sigma}_x=|1\rangle\langle 2|+|2\rangle\langle 1|$. The LD parameter is defined in terms of the ionic mass $m$ and laser wave number $k$ as $\eta=k\sqrt{m\omega/2}$, and the relative position of the trap in the standing wave laser field is determined by $\phi$.  

In order to gain a deeper understanding of the evolution we discuss
the approximations regularly applied in trapped ion models. These
often result in analytically solvable effective Hamiltonians. In the
LD regime, $\eta\ll1$, we may expand the cosine function to obtain
\begin{equation}\label{hamexp}
H\!\approx\!\omega\hat{a}^\dagger\hat{a}+\!\frac{\Delta}{2}\hat{\sigma}_z+
\lambda\Big[\cos(\phi)-\eta\sin(\phi)(\hat{a}^\dagger+\hat{a})\Big]
\hat{\sigma}_x+\mathcal{O}(\eta^2).
\end{equation}
In this regime, the effect of the phase shift $\phi$ is clear. For
$\phi=k\pi$, where $k$ is an integer, the ionic vibrational states
are not coupled by the external laser  ({\it carrier pumping}),
whereas if $k=1/2,3/2,\ \ldots$ the vibrational states do couple
({\it sideband pumping}) \cite{com2}. In the latter case, the
Hamiltonian (\ref{hamexp}) coincides with the Rabi model
\cite{oldwine}. Neglecting the virtual processes (i.e.,
$\hat{a}\hat{\sigma}^-$ and $\hat{a}^\dagger\hat{\sigma}^+$
corresponding to simultaneous deexcitation/excitation of the field
and the ion) one derives the regular JC Hamiltonian in the RWA
\begin{equation}
H_{JC}=\omega\hat{a}^\dagger\hat{a}+\frac{\Delta}{2}\hat{\sigma}_z+
g\left(\hat{a}^\dagger\hat{\sigma}^-+\hat{\sigma}^+\hat{a}\right),
\end{equation}
where in this case $g=\lambda\eta$ and
$\hat{\sigma}^+=|2\rangle\langle 1|$ and
$\hat{\sigma}^-=|1\rangle\langle 2|$. The JC model is analytically
solvable and the eigenvalues of its corresponding Hamiltonian are known as (up to an overall
constant)
\begin{equation}\label{jceig}
E_{JC}^\pm(n)=\omega n\pm\sqrt{\frac{(\Delta-\omega)^2}{4}+g^2n},\hspace{1cm}n=0,1,2,\ \ldots~.
\end{equation}

As pointed out, the effective RWA Hamiltonian, in principle, is only
justified for $\cos(\phi)=0$, whereas in general situations more
sophisticated RWA approaches must be considered. Turning to an
interaction picture with respect to the free field and ion,
Hamiltonian (\ref{hamexp}) becomes
\begin{equation}
\begin{array}{lll}
H_I &  = & U_I H U_I^{-1} =\\ \\
& = & \lambda\cos(\phi)\left(\hat{\sigma}^+\mathrm{e}^{-i\Delta t}+\hat{\sigma}^-\mathrm{e}^{i\Delta t}\right)\\ \\ & & -\eta\lambda\sin(\phi)\left(\hat{a}^\dagger\hat{\sigma}^-\mathrm{e}^{-i(\omega-\Delta)t}+\hat{\sigma}^+\hat{a}\mathrm{e}^{i(\omega-\Delta)t}\right)\\ \\
& & -\eta\lambda\sin(\phi)\left(\hat{a}^\dagger\hat{\sigma}^-\mathrm{e}^{-i(\omega+\Delta)t}+\hat{\sigma}^+\hat{a}\mathrm{e}^{i(\omega+\Delta)t}\right).
\end{array}
\end{equation}
where
$U_I=\exp{(-i\omega\hat{a}^\dagger\hat{a}t)}\exp{(-i\frac{\Delta}{2}\hat{\sigma}_zt)}$.
This introduces three different time scales of the interaction,
suggesting an application of the RWA. It is known, however, that the
validity of the RWA does not solely depend on the involved time
scales, but also on the relative amplitudes of the coupling
parameters \cite{oldwine,jonaswp}. In particular, only for ion-field
couplings much smaller than the ion level separation and vibrational
spacing �-- $\lambda\eta\sin(\phi)\ll\Omega,\omega$ and
$\lambda\cos(\phi)\ll\Omega,\omega$, �-- is the RWA justified. We
then note: Firstly, the Lamb-Dicke regime seems to favor application
of the RWA, which, in fact, we have verified by numerical
calculation. Secondly, the angle $\phi$ is of importance for the
validity of the RWA and together with the three characteristic time
scales it becomes clear that a straight-forward application of the
RWA in the basis presented above is non-trivial. The first
observation will be of importance further on when we introduce the
adiabatic approximation. The second is indeed an interesting issue
that was resolved by Wu and Yang \cite{rwasw}. Their idea is that
also the third term of the expanded Hamiltonian (\ref{hamexp}) is
considered as a ``free'' part, and these first three terms define
the interaction picture and the proper time scales. Within this
frame, the slowly oscillating terms describe a JC type of
interaction between the field (vibrations) and ``dressed'' atomic
states \cite{com1} differing from the $|1\rangle$ and $|2\rangle$
states.

Yet another, but related, approach is to use the expansion
\cite{mankov,noLD}
\begin{equation}
\begin{array}{lll}
\displaystyle{\cos[\eta(\hat{a}^\dagger+\hat{a})+\phi]}\! & =\! &
\displaystyle{\mathrm{e}^{-\eta^2/2}\sum_{k=0}^\infty\epsilon_k
\frac{(-\eta)^k}{k!}\cos\left(\phi+k\frac{\pi}{2}\right)}\\ \\ & &
\!\!�\!\left[\hat{a}^kf_k(\hat{n}+k)¡\!+\!f_k(\hat{n}+k)(\hat{a}^\dagger)^k\right]\!,
\end{array}
\end{equation}
where
\begin{equation}
f_k(\hat{n})=\sum_{m=0}^\infty\frac{(\hat{n}-m+1)_m}{(k+1)_mm!}=k!\frac{L_{\hat{n}}^k(\eta^2)}{(\hat{n}+1)_k}
\end{equation}
with $(\hat{n}-m+1)_m=\hat{n}(\hat{n}-1)\hdots(\hat{n}-m+1)$,
$\hat{n}=\hat{a}^\dagger\hat{a}$, $\epsilon_k=1/2$ for $k=0$ and
$\epsilon_k=1$ otherwise and $L_{\hat{n}}^k(\eta^2)$ are the
generalized Laguerre polynomials. Imposing suitable resonance
conditions ($|\Delta-\omega|\ll|\Delta-k\omega|$, for $k=0,2,3,4,\
\ldots$) between the constituting frequencies, we may for proper
amplitudes of the couplings employ a RWA to regain a JC like
interaction in terms of ``deformed'' oscillator operators
$\hat{A}=\hat{a}f_1(\hat{n}+1)$. The $\hat{A}$ and $\hat{A}^\dagger$
define generalized boson operators obeying modified boson algebras
and they specify, for instance, non-linear Fock and coherent states
being eigenstates of $\hat{A}^\dagger\hat{A}$ and $\hat{A}$,
respectively. Note that, in some sense, the new ladder operators can
be viewed as describing a non-linear oscillator with
amplitude-dependent frequency. Additionally, it is known that the
anharmonicity of an oscillator greatly affects physical phenomena
like collapse and revivals \cite{wprev}. Related algebraic
techniques, in particular shape invariance and SUSY, can therefore
be a useful tool for describing molecular dynamics where oscillators
are in general anharmonic \cite{susy}. We remark, however, that most
of these models \cite{susy} employ the standard RWA as it is used in
the JC model, while it is clear that this is in general less
justified than for the JC model, due to varying distance between
energy eigenvalues, for instance.

We pointed out above that the relative size between the effective
laser-ion coupling $\lambda$ and the two-level spacing $\Omega$ is
crucial for the validity of the RWA. Also the relation between
$\lambda$ and $\omega$ is important. In general, for moderate phonon numbers ($n\leq10$), the
strong-excitation regime \cite{strong}, $\lambda\gg\omega$,
invalidates the RWA and the LD approximations, while the opposite
weak-excitation regime \cite{low} favors them. It should be
emphasized, though, that, eventually, the applicability of the
approximations depends on all parameters, including $\Omega$ and
$\eta$, and their mutual relations.

Now, rather than by the boson creation and annihilation operators as
above, we represent the Hamiltonian in terms of conjugate variables
\begin{equation}
\begin{array}{l}
\displaystyle{\hat{x}=\sqrt{\frac{1}{2m\omega}}\left(\hat{a}^\dagger+\hat{a}\right)},\\ \\
\displaystyle{\hat{p}=i\sqrt{\frac{m\omega}{2}}\left(\hat{a}^\dagger-\hat{a}\right)},
\end{array}
\end{equation}
obeying the regular canonical commutator relation. In this nomenclature we find
\begin{equation}\label{ham3}
H=\frac{\hat{p}^2}{2m}+\frac{m\omega^2}{2}\hat{x}^2+\left[\begin{array}{cc}
                                                           \displaystyle{\frac{\Delta}{2}} & \lambda\cos(k\hat{x}+\phi) \\
                        \lambda\cos(k\hat{x}+\phi) & -\displaystyle{\frac{\Delta}{2}}
                                                          \end{array}\right],
\end{equation}
where $k$ is the external laser wave number and $m$ the mass of the ion.
This representation of the Hamiltonian serves as the starting point
of our analysis; for a given initial state, its wave packet will
evolve on two coupled harmonic oscillators according to eq.
(\ref{ham3}). We note that in $x$-representation a \emph{Fock state}
of the vibrations, $\hat{n}|n\rangle=n|n\rangle$, or a
\emph{coherent state}, $\hat{a}|\alpha\rangle=\alpha|\alpha\rangle$,
read
\begin{equation}
\begin{array}{l}
 \displaystyle{\varphi_n(x)=\langle x|n\rangle=\left(
 \frac{\sqrt{m\omega}}{2^n\sqrt{\pi}n!}\right)^{1/2}
 H_n(\sqrt{m\omega}x)\mathrm{e}^{-\frac{m\omega x^2}{2}}}, \\ \\
\displaystyle{\varphi_\alpha(x)=\langle x|\alpha\rangle=
\left(\frac{m\omega}{\pi}\right)^{1/4}\mathrm{e}^{-i[\Im(\alpha)]^2}
\mathrm{e}^{-\frac{m\omega(x-\sqrt{\frac{2}{m\omega}}\alpha)^2}{2}}}.
\end{array}
\end{equation}
Here, $H_n(x)$ is the $n$:th Hermite polynomial.

\subsection{Base representations}\label{ssec2b}
To gain a deeper intuition of the problem it is in order to discuss
various bases that will be used in the following. One natural representation
is that of \emph{bare basis states}, which are the eigenstates of the free
Hamiltonian part $\{|n,1\rangle,|n,2\rangle\}$, where $n$ represents the
$n$:th Fock eigenstate of the harmonic oscillator. In $x$-representation
the bare basis states are written as $\Bigg\{\langle x |n,1\rangle = \varphi_n(x)
\left[ \begin{array}{c}1 \\ 0\end{array}\right],\langle x |n,2\rangle = \varphi_n(x)\left[\begin{array}{c}0 \\ 1\end{array}\right]\Bigg\}$.\\
A general \emph{bare state} is given by $\psi_1(x)|1\rangle$ or
$\psi_2(x)|2\rangle$ for some normalized wave function $\psi_i(x)$.
From the form on which the Hamiltonian (\ref{ham3}) is presented, it
is clear that the external laser field enters in the off-diagonal
terms and hence couples these states. The bare states can be used to
define \emph{bare potential curves} as the diagonal potential
elements of the Hamiltonian in the ionic basis $|1\rangle$ and
$|2\rangle$. Thus, the bare potentials are two centered oscillators
shifted in energy with either $\Delta/2$ or $-\Delta/2$.

The eigenstates of the full Hamiltonian are called \emph{dressed
basis states} and denoted by $|\chi_n^+\rangle$ and
$|\chi_n^-\rangle$. We have introduced the superscript $\pm$ since,
for a wide range of parameters, the eigenstates come in pairs, which
is true in particular when the RWA or the LD approximation or both
have been imposed, as seen in (\ref{jceig}) for the JC model. In
these regimes, not necessarily only for the JC model, ``simple''
analytic solutions to the problem are available, whereas in the
general case this is not true. Note that there is no obvious
corresponding set of dressed potential curves.

The above two bases are part of the the conventional terminology of
quantum optics. We now turn to a more convenient representation of
coupled molecular electronic states. For this purpose, we employ the
$x$-representation and rotate the Hamiltonian (\ref{ham3}) around
the $(\hat{\sigma}_x+\hat{\sigma}_z)$-axis by the unitary operator
$U_1=\frac{1}{\sqrt{2}}(\hat{\sigma}_x+\hat{\sigma}_z)$ giving
\begin{equation}
\begin{array}{lll}
\tilde{H}_1 & = & U_1HU_1^{-1}\\ \\
&\! =\! & \displaystyle{\!\!\frac{\hat{p}^2}{2m}\!+\!
\frac{m\omega^2\!}{2}\hat{x}^2}\!+\!\left[\begin{array}{cc}\!\!
\lambda\cos(k\hat{x}+\phi) & \displaystyle{\frac{\Delta}{2}}\\
\displaystyle{\frac{\Delta}{2}} & \!\!-\lambda\cos(k\hat{x}+\phi)\!                                                         \end{array}\right]\!.
\end{array}\label{ham4}
\end{equation}
This transformation swaps the off-diagonal terms with the diagonal
ones of the last term in the Hamiltonian. The form of $\tilde{H}_1$
has many similarities with the Hamiltonian describing a diatomic
molecule in a diabatic representation. The off-diagonal elements in
(\ref{ham4}) are in potential form, i.e., they do not contain
differential operators. Here, the couplings are
$\hat{x}$-independent, which however is also frequently assumed in
models used in molecular physics, e.g., the Landau-Zener model
\cite{lz}. Note that in our ion-trap system $\hat{x}$ represents the
spatial center-of-mass position of the ion in the trap, whereas in
the molecular counterpart $\hat{x}$ is the internuclear distance.
The operator $U_1$ transforms the bare internal ionic states as
\begin{equation}
\begin{array}{l}
\displaystyle{|-\rangle=U_1|1\rangle=\frac{1}{\sqrt{2}}\left[\begin{array}{c}1
\\ 1\end{array}\right]},\\ \\
\displaystyle{|+\rangle=U_1|2\rangle=\frac{1}{\sqrt{2}}\left[\begin{array}{c}1
\\ -1\end{array}\right],}
\end{array}
\end{equation}
which defines the \emph{diabatic states} as $\langle
x|\psi\rangle|\pm \rangle = \psi_\pm(x)|\pm\rangle$, and the
corresponding \emph{diabatic potential curves} as the diagonal
potential matrix terms of (\ref{ham4}). Thus, contrary to the bare
curves, these are centered oscillators modified by
$\pm\lambda\cos(k\hat{x}+\phi)$ and depending on the parameter
amplitudes very different types of diabatic potentials may be
obtained. For example, if the oscillating cosine function dominates
over the harmonic oscillator (in the spatial range of interest), the
system is semi-periodic and is best understood from Bloch or Floquet
theory \cite{jonas1}. In the opposite limit of a small $\lambda$ and
$k$ we may expand the cosine as in (\ref{hamexp}) to find the two
centered oscillators both shifted and displaced according to the
added terms
$\pm\lambda\left[\cos(\phi)-\eta\sin(\phi)\hat{x}\right]$. In this
paper, we are interested in a regime intermediate between the two.

Given the Hamiltonian in the form of (\ref{ham3}), the transformation
\begin{equation}
U_2=\left[\begin{array}{cc}\cos(\theta) & -\sin(\theta) \\
           \sin(\theta) & \cos(\theta)
          \end{array}\right]
\end{equation}
with
\begin{equation}
\tan(2\theta)=\frac{2\lambda\cos(k\hat{x}+\phi)}{\Delta}
\end{equation}
diagonalizes the last term of the Hamiltonian. However, since $U_2$
is $\hat{x}$-dependent it does not commute with $\hat{p}$ and
explicitly we get
\begin{equation}\label{adtrans}
\begin{array}{l}
\tilde{H}_2= \displaystyle{U_2HU_2^{-1}=\frac{\hat{p}^2}{2m}+
\frac{m\omega^2}{2}\hat{x}^2+(\partial\theta)^2}\\ \\
+\left[\begin{array}{cc}\varepsilon(\hat{x}) &
\partial^2\theta-2i(\partial\theta)\hat{p} \\
-\partial^2\theta+2i(\partial\theta)\hat{p} & -\varepsilon(\hat{x})
\end{array}\right],
\end{array}
\end{equation}
where $\partial f\equiv\partial f/\partial x$ and
\begin{equation}
\varepsilon(\hat{x})=\sqrt{\Delta^2/4+\lambda^2\cos^2(k\hat{x}+\phi)}.
\end{equation}
These diagonal terms define the \emph{adiabatic potential curves},
while $U_2$ acting on the bare states gives the \emph{adiabatic
states}; $\langle x|\psi\rangle|i \rangle_{ad}=
\psi(x)|i\rangle_{ad}=\psi(x)U_2|i\rangle$, for $i=1,2$. Note that
$|i\rangle_{ad}$ is in general $x$-dependent. The diagonal and
non-diagonal terms containing $\partial\theta$ and
$\partial^2\theta$ are the non-adiabatic corrections, and if these
play a minor role in the wave packet evolution, the dynamics is said
to be adiabatic. For the present system, we have
\begin{equation}\label{noad}
\begin{array}{l}
\displaystyle{\partial\theta=-\frac{\Delta k\lambda s}{\Delta^2+\lambda2c^2}},\\ \\
\displaystyle{\partial^2\theta=-\frac{\Delta\left[\lambda k^2c(\Delta^2+4\lambda^2c^2)-8\lambda^3k^2cs^2\right]}{\left(\Delta^2+4\lambda^2c^2\right)^2}},
\end{array}
\end{equation}
where $s=\sin(k\hat{x}+\phi)$ and $c=\cos(k\hat{x}+\phi)$, and
typically the second, higher derivative term, is smaller. For
diatomic molecules an equivalent transformation from the diabatic to
an adiabatic representation can be applied. Consequently, also then
non-adiabatic couplings are both $\hat{x}$- and $\hat{p}$-dependent.
We remind, as discussed in the previous section, that the RWA and LD
approximations are highly related in terms of validity ranges and
especially $\eta\ll1$ (or likewise for small $k$) is a constraint on
their implementation. Here, as well, we note that for small $k$ the
non-adiabatic coupling terms (\ref{noad}) become small. On top, we
claimed that $\lambda<\Delta$ is also a condition for the
applicability of the RWA and the LD approximation, which from
(\ref{noad}) again is seen to favor the adiabaticity. We have
verified by numerical calculation that all three approximations more
or less share the same parameter validity range. From (\ref{noad})
one notes that also the case when $\Delta\equiv0$ gives vanishing
non-adiabatic corrections. This limit, however, does not represent a
proper adiabatic one, but rather the diabatic limit, and is hence
sometimes referred to as the anti-adiabatic limit. Since we consider
parameters where the adiabatic approximation breaks down, it follows
that we can apply neither the RWA nor the LD approximation.

\begin{figure}[!ht]
\begin{center}
\includegraphics[width=7cm]{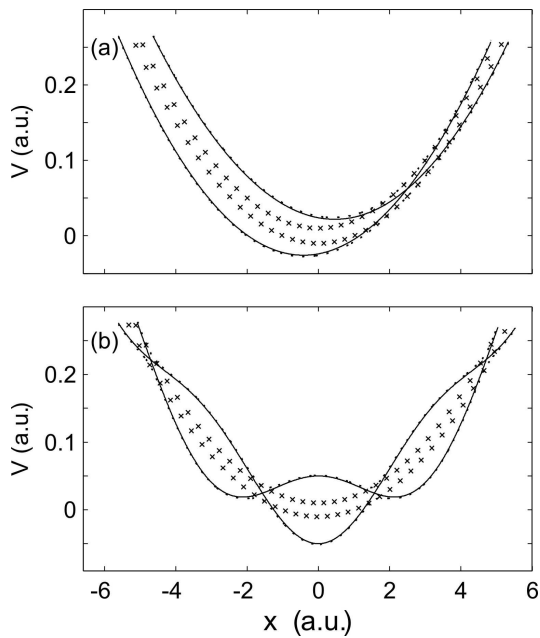}
\vspace{0.4cm} \caption{ Example potentials: bare (crosses),
diabatic (solid line) and adiabatic (dots). In (a), the potentials
only possess one crossing in the shown interval, while (b) displays
an example of several crossings. The parameters (in atomic units)
are: $m=80000$, $\omega=0.0005$, $\lambda=0.05$, $\Delta=0.02514$ in
both plots, while $k=0.2$ and $\phi=1.07249074$ rad in (a) $k=1$ and
$\phi=0$ rad in (b). } \label{fig2}
\end{center}
\end{figure}

In Table I, we summarize the various bases, while in Fig.~\ref{fig2}
we show some typical examples of the corresponding potentials. Note
that at the crossing point, when the cosine function is zero, the
bare and adiabatic potentials coincide, while away from the crossing
the diabatic and adiabatic curves approach each other for the
current set of parameters. We may point out that different definitions of the above bases exist, and in particular in quantum optics is the bare-diabatic and the dressed-adiabatic representations often 
identical. Identifying bare states as diabatic ones and dressed states with adiabatic ones does not, however, properly coinside with the common definitions in molecular and chemical physics. Adiabatic states are in general not eigenstates of the full Hamiltonian, as dressed states are, and diabatic states usually approximate the adiabatic ones far from curve crossing regions, which is not the case of bare states.

\begin{widetext}
\begin{center}
\begin{table}\label{tabe1}
\begin{tabular}{|c|c|c|}\hline
State & Definition & Potentials \\ \hline\hline
Bare & $\psi_i(x)|i\rangle$ & $V_\pm^B(\hat{x})=\frac{m\omega^2}{2}\hat{x}^2\pm
\frac{\Delta}{2}$ \\ \hline
Dressed & Eigenstates of full Hamiltonian (\ref{ham3}), $|\chi_n^\pm
\rangle$ & - \\ \hline
Diabatic &  $\psi_-(x)|-\rangle=\psi_-(x)U_1|1\rangle,\,\psi_+(x)|+
\rangle=\psi_+(x)U_1|2\rangle$ & $V_\pm^D(\hat{x})=\frac{m\omega^2}{2}
\hat{x}^2\pm\lambda\cos(k\hat{x}+\phi)$ \\ \hline
Adiabatic & $\psi_i(x)|i\rangle_{ad}=\psi_i(x)U_2|i\rangle$ & $V_\pm^A(\hat{x})=
\frac{m\omega^2}{2}\hat{x}^2\pm\varepsilon(\hat{x})$ \\ \hline
\end{tabular}
\caption{States and potentials used in this article. Here, $i=1,2$
and $\psi_i(x)$ is a general normalized wave function.}
\end{table}
\end{center}
\end{widetext}

\section{Dynamics -- bistable motion}\label{sec3}
The existence of bistable evolution is a direct consequence of
interference between the constitutent wave packet parts, as shown in
Fig. \ref{fig1}. The bistable motion turns out to be stable over
very long time periods, showing clear oscillations, collapses and
revivals \cite{dong}. In order for the wave packet fractions to
overlap and interfere correctly, a necessary, but not sufficient,
condition for bistable motion is that the two parts of the splited
wave packet should return simultaneously to the starting point. We
denote this period of time by $T_{cl}$, where the index $cl$ stands
for classical, and it is determined from the first maximum of the
autocorrelation function, as will be described below. Further, let
$P_i(t)$ be the population in diabatic state $|i\rangle$ (we hence
trace out the field degrees of freedom). If, initially, the wave
packet is in the diabatic state $i$ we define
$P_{sp}=1-P_i(t=T_{cl}/2)$, which gives the population in the
opposite diabatic state at half a classical period. In other words,
$P_{sp}$ in a sense measures the amount of splitting of the wave
packet as it traverses the crossing and, hence, for times shorter
than the collapse time $P_i(jT_{cl})\approx1$ ($j=0,1,2,\ \ldots$)
for a bistable trajectory. Naturally, $0\leq P_{sp}\leq1$, where
$P_{sp}=0$ corresponds to diabatic and $P_{sp}=1$ to adiabatic
evolution, and in particular, bistable trajectories are not
restricted to a certain $P_{sp}$. It is clear, however, that for the
non-trivial intermediate values of $P_{sp}$, these types of
interferences are sensitive to parameter values. In addition to
bistable trajectories, the parameters may be tuned so that
$P_i((2j+1)T_{cl})\approx0$ and $P_i(2jT_{cl})\approx 1$, where
again $j=0,1,2,\ \ldots$, which defines an \emph{astable
trajectory}. Here, however, we focus on the more stable bistable
case.

For a general set of bound coupled (or uncoupled) potential curves
we may expand the eigenenergies around the mean set of quantum
numbers $n_0$ forming the wave packet. For a single discrete number
$n$ this reads
\begin{equation}\label{eigexp}
\begin{array}{lll}
E(n) & \approx & \displaystyle{ E(n_0)+E'(n_0)(n-n_0)+\frac{E''(n_0)}{2}(n-n_0)^2} \\ \\
& & \displaystyle{+\frac{E'''(n_0)}{6}(n-n_0)^3}+\ \ldots~,
\end{array}
\end{equation}
where $E'(n_0)=(dE(n)/dn)_{n=n_0}$ and so forth, and $n_0\gg1$ is
the ``average'' quantum number of the wave packet.  The wave packet
is assumed to be composed of eigenstates with quantum numbers $n$
fairly localized around $n_0$. We define the different time scales
\begin{equation}\label{timescales}
\begin{array}{lll}
T_{cl}=\frac{2\pi}{|E'(n_0)|}, & T_{rev}=\frac{2\pi}{|E''(n_0)|/2}, &
T_{sup}=\frac{2\pi}{|E'''(n_0)|/6},
\end{array}
\end{equation}
characterizing the \emph{classical vibration}, \emph{revival} and
\emph{superrevival} time, respectively. Note that for a single
harmonic oscillator $T_{cl}=2\pi/\omega$ and the higher order terms
vanish, which is equivalent to letting the latter characteristic
time scales go to infinity. For two uncoupled different harmonic
oscillators 1 and 2 one has the respective classical periods
$T_{cl}^{(1)}=2\pi/\omega_1$ and $T_{cl}^{(2)}=2\pi/\omega_2$, where
$\omega_i$ is the frequency of the oscillator $i$. The combined
classical period becomes $T_{cl}=T_{cl}^{(1)}k=T_{cl}^{(2)}l$, where
$k$ and $l$ are the smallest possible integers obeying the condition
$k\omega_2=l\omega_1$. Thus, it is clear that for multi-level
systems the typical time scales can be very long. As an example,
returning to the JC model (\ref{jceig}), the revival time is usually
defined in the quantum optics community as the time it takes for the
constituent neighboring Rabi frequencies $\Omega_n=\sqrt{(
 \Delta-\omega)^2/4+g^2n}$ to differ by $\pi$,
\begin{equation}\label{jcrev}
\left(\Omega_{n_0+1}-\Omega_{n_0}\right)T_{rev}=\pi.
\end{equation}
Consequently, the JC revival time as it is given in (\ref{jcrev}) is
more reminiscent of the classical period rather than the revival one
according to the definition (\ref{timescales}). Also the fractional
and superrevivals as discussed in the literature of the JC model
differ from the ones of eq.~(\ref{timescales}), see \cite{jcsup}.

The wave packet propagation is carried out using the Chebychev
polynomial method \cite{cheby}. We assume the initial state to be of
the form $\psi(x,0)|+\rangle$, where $\psi(x,0)$ is taken to be a
minimum uncertainty Gaussian function
\begin{equation}
\Psi(x,0)=\psi(x,0)|+\rangle=\left(\frac{1}{2\pi\sigma^2}\right)^{1/4}
\mathrm{e}^{-\frac{(x-x_0)^2}{4\sigma^2}},
\end{equation}
including also the case of coherent states. This wave packet is let
to evolve over long time periods, typically $T_{rev}$. The
quantities of interest for us are
\begin{equation}\label{quant}
\begin{array}{lll}
W(t)=\langle\hat{\sigma}_z\rangle, & &\mathrm{\emph{Inversion}}\\ \\
A(t)=\displaystyle{\int \Psi^*(x,t)\Psi(x,0)dx},& &
\mathrm{\emph{Autocorrelation}}\\ \\
S_I(t)=-\mathrm{Tr}\big[\rho_I\log(\rho_I)\big],& & \mathrm{\emph{Entropy}}
\end{array}
\end{equation}
where $\Psi(x,t)$ is the full two-level wave packet at time $t$, and
$\rho_I=\mathrm{Tr}_F\big[\rho\big]$ is the reduced density operator
for the ion, obtained once the field degrees of freedom of the full
density operator $\rho=|\Psi(x,t)\rangle\langle\Psi(x,t)|$ have been
traced out. Note that $W(t)$ is the inversion between the bare
states $|2\rangle$ and $|1\rangle$. The autocorrelation function
determines the overlap between the initial state with the one at
time $t$, while the \emph{von Neumann entropy} is a measure of the
degree of entanglement shared between the field and the ion. For
initial pure states, as in our case, the entropy is the same for the
field and the ion; $S_I(t)=S_F(t)$ \cite{araki}. An advantage of our
numerical wave packet approach is that, once the full wave packet
$\Psi(x,t)$ is given, all the above quantities are easily
calculated, as well as other relevant properties.

\subsection{Short time evolution}\label{ssec3a}
By short times we refer to time scales of order $T_{cl} \ll T_{rev}$
corresponding to a single or few oscillations of the total wave
packet. We tune the parameters to have 50-50 splitting of the wave
packet at the curve crossing, i.e. $P_{sp} = 1/2$, corresponding to
maximal entanglement. The initial wave packet is localized around
$\alpha=x_0\sqrt{m\omega}/\sqrt{2}=6\sqrt{m\omega}/\sqrt{2}\approx26.8$
for the considered mass, and thus for the chosen parameters the
motion is highly excited with a mean phonon vibrational number
$n_0=720$. Its width is either $\sigma=0.047$ (highly squeezed
\cite{sq}), or we pick $\sigma$ as for a coherent state with given
mass $m$. In atomic units $m=80000$, which corresponds to $42$ amu.
The results are applicable to any mass, however, provided the time
is scaled accordingly. The rest of the parameters are as in
fig.~\ref{fig2} (a), except $\Delta$ that should be divided by 5.

\begin{figure}[!ht]
\begin{center}
\includegraphics[width=7cm]{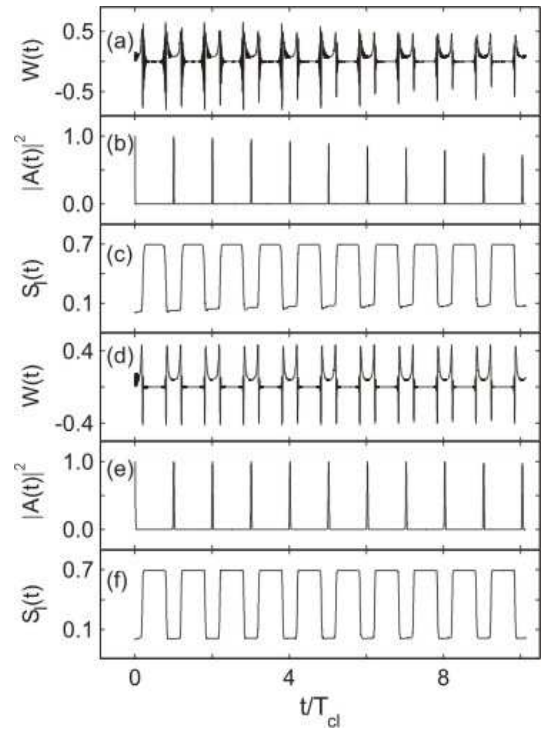}
\caption{ Inversion $W(t)$, autocorrelation $A(t)$ and entropy
$S_I(t)$ for a coherent (a)-(c) and a squeezed (d)-(f) initial
state. See text for the parameters.} \label{fig3}
\end{center}
\end{figure}

In fig.~\ref{fig3} we display the numerical results of the
quantities (\ref{quant}), (a)-(c) refer to an initial coherent state
and (d)-(f) an initial Gaussian distribution with $\sigma=0.047$. We
note that in both cases almost perfect 50-50 splitting is obtained
and thus the system is maximally entangled on one side of the level
crossing, while it is disentangled on the opposite side. The state can be written as
\begin{equation}
\begin{array}{lll}
\Psi(x,t) & = & \psi_+(x,t)|+\rangle+\psi_-(x,t)|-\rangle\\ \\
& = & \displaystyle{\frac{1}{\sqrt{2}}\left(\psi_-(x,t)+\psi_+(x,t)\right)|1\rangle}\\ \\
& & \displaystyle{+\frac{1}{\sqrt{2}}\left(\psi_-(x,t)-\psi_+(x,t)\right)|2\rangle}.
\end{array}
\end{equation}
For 50-50 splitting, and well separated splitted wave packets, we have that the vibrational wave functions of the bare states $|1\rangle$ and $|2\rangle$ are orthogonal, in agreement with the maximal entanglement. At this instant, the combined bi-partit ion-phonon system is in an $EPR$ Schr\"odinger cat state. To a first approximation, we may assume the wave packets to keep their shape over a time period of one classical oscillation. That is, for initial coherent states, the splitted states are still coherent with modulated phase and amplitude. This is indeed verified below, where we considers the dynamics in phase space. Thus, the phonon distrubutions in this case are Poisonians for the diabatic states, while for the bare states they would be made out of the sum of two coherent states with equal phase but different amplitude. 

Calculating the variance of the wave packets, one notes that the squeezed state
\cite{sq}, $\sigma=0.047$, shows typical breathing motion while
traversing the potential back and forth. From the decrease in
amplitudes, the quantities corresponding to the coherent state is
seen to be more sensitive than for the squeezed state, or in other
words, the collapse time is faster. This fast dephasing comes about
due to the larger spread in phonon numbers $n$ for the coherent
initial state.  It also shows up as rapid oscillations in the
inversion around the curve crossings.

The plots also give a measure of the characteristic transition time
at the level crossing \cite{vitanov1}, and one notes that this time
is rather short compared to the full time $T_{cl}$, which is indeed
known from other similar models \cite{jonas3}. Interestingly, we
have found that short crossing time as well as the sharp peaks with
large amplitudes in the inversions, (a) and (d), around the
crossings are typical for the bistable cases. The bare state
inversion can be expressed in terms of the diabatic wave packets as
$\langle\hat{\sigma}_z\rangle=2\Re\left[\int\psi_+^*(x,t)\psi_-(x,t)dx\right]$,
and is hence related to the coherence of diabatic wave packets. We
note that the total energy can be written as
\begin{equation}
E_{tot}(t)=\sum_{i=+,-}\int\psi_i^*(x,t)H_{ii}\psi_i(x,t)dx+\frac{\Delta}{2}W(t),
\end{equation}
where $H_{++}$ and $H_{--}$ are the two diagonal matrix elements of
the Hamiltonian (\ref{ham4}). From this it is seen that the bare
state inversion is a measure of the diabatic energy transfer.

At first, it may seem surprising that the inversion $W(t)$ can be
non-zero at the right side of the crossing, even if the
autocorrelation $A(t)$ is close to 1 and the entropy $S_I(t)$ is
close to zero. The zero entropy, however, is not a contradiction as
the field and ion states can be separated, provided the constituent
wave packets fractions $\psi_+(x,t)$ and $\psi_-(x,t)$ differ merely
by a complex constant. For the inversion we have
\begin{equation}
\begin{array}{lll}
W(t) & = & \langle\psi_2(t)|\psi_2(t)\rangle-\langle\psi_1(t)|\psi_1(t)\rangle \\ \\
& = & 2\Re\left[\langle\psi_+(t)|\psi_-(t)\rangle\right]\\ \\
& \leq & 2\sqrt{\left[1-\langle\psi_-(t)|\psi_-(t)\rangle\right]\langle\psi_-(t)
|\psi_-(t)\rangle},
\end{array}
\end{equation}
while for the autocorrelation
\begin{equation}
\begin{array}{lll}
|A(t)|^2 & = & |\langle\Psi(t)|\Psi(0)\rangle|^2= |\langle\psi_+(t)|\psi_+(0)\rangle|^2\\ \\
& \leq & 1-\langle\psi_-(t)|\psi_-(t)\rangle.
\end{array}
\end{equation}
For sufficiently small $\langle\psi_-(t)|\psi_-(t)\rangle$ it
follows that $A(t)$ can be close to one even if $W(t)$ is not
exactly zero.

\begin{figure}[!ht]
\begin{center}
\includegraphics[width=8cm]{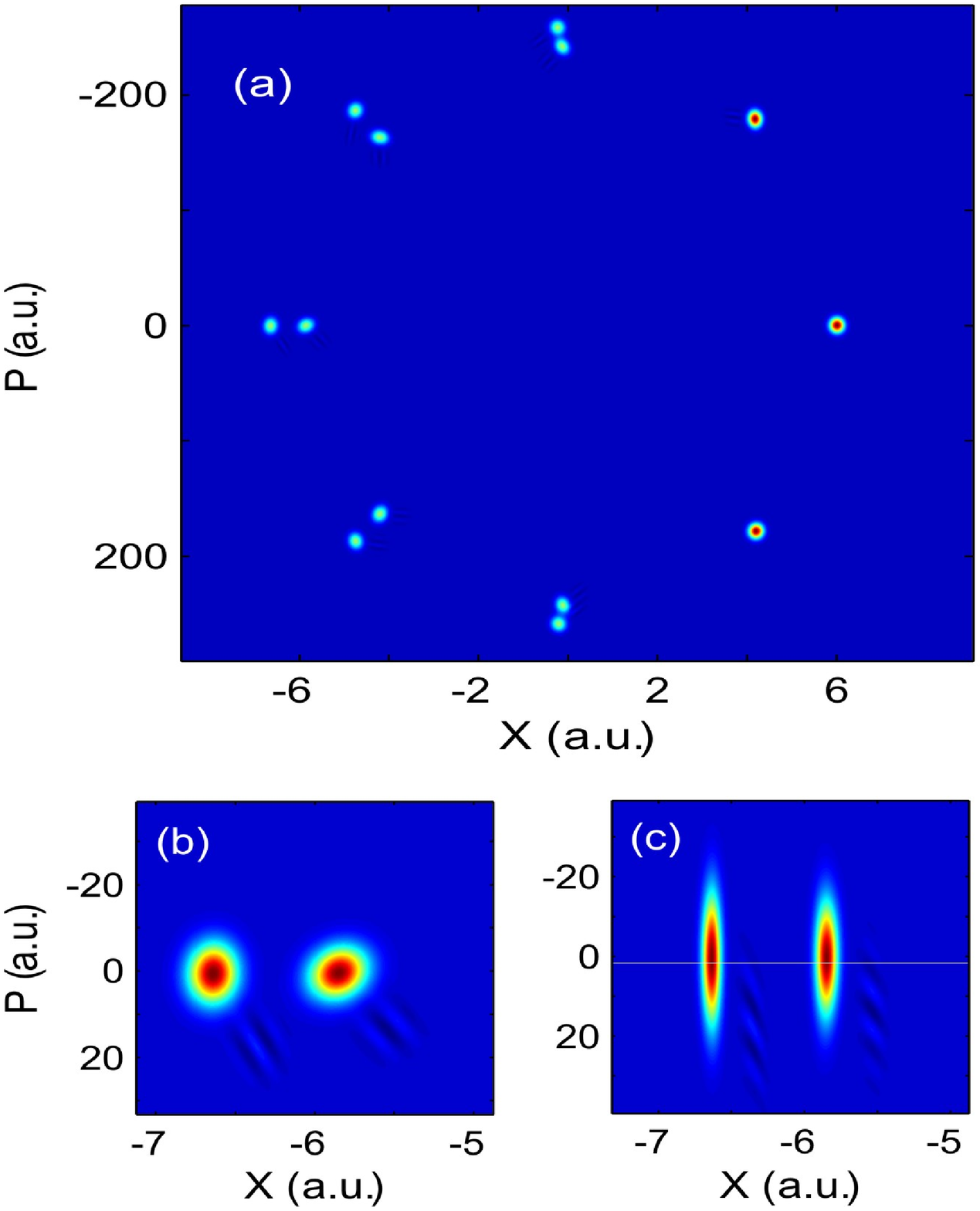}
\includegraphics[width=8cm]{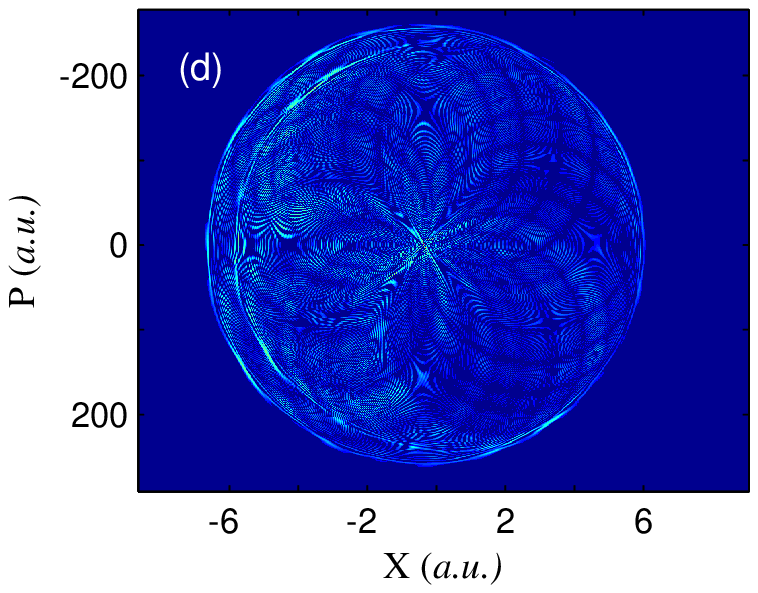}
\caption{ (Colour online) The Wigner distribution for an initial
coherent state (a) and (b) and for a squeezed state (c)  and (d).
The first plot shows snap-shots throughout one period of
oscillation, while (b) and (c) zooms in at the distributions around
half oscillation time. In (d) the distribution at the time of
collapse is displayed.} \label{fig4}
\end{center}
\end{figure}

The wave packet splitting manifests itself also in the phase space
distributions \cite{oldwine}. In Fig.~\ref{fig4} we display
snap-shots of the Wigner distribution \cite{wig}
\begin{equation}\label{wigner}
\begin{array}{lll}
\mathcal{W}(p,x) & = & \displaystyle{\frac{1}{(2\pi)^2}\int d\xi\int d\tau\,
\mathrm{e}^{i(\tau p+\xi x)}\mathrm{Tr}_I\Big[\mathrm{e}^{-i(\tau p+\xi x)}\rho\Big]}\\ \\
& = & \displaystyle{\frac{1}{\pi}\int dy\,\mathrm{e}^{-i2py}\Big[\psi_+^*(x-y)\psi_+(x+y)}\\ \\
& & +\psi_-^*(x-y)\psi_-(x+y)\Big]
\end{array}
\end{equation}
over one period of oscillation for a coherent state (a) and a zoom
in at the Wigner distribution close to half an oscillation for a
coherent (b) and squeezed state (c). In (d) we show the Wigner
distribution after the collapse at approximately
$t\approx2000T_{cl}$ for a squeezed state. The spreading of the
distribution over the whole accessible phase-space is a character of
the collapse. At times of fractional revivals, the wave packet forms
multiple localized bumps \cite{bumprev}. The two constitute parts of
$\mathcal{W}(p,x)$ approximately follow the ``classical'' phase
space trajectories
\begin{equation}
\frac{\partial x_c}{\partial t}=\frac{\partial H_�^{(i)}}{\partial p_c},
\hspace{1cm}\frac{\partial p_c}{\partial t}=-\frac{\partial H_�^{(i)}}{\partial x_c},
\end{equation}
with the adiabatic and diabatic Hamiltonians
$H_�^{(i)}=\frac{p_c^2}{2m}+V_�^{(i)}(x)$ and $i=A,D$. The small
anharmonicity of the potentials is reflected in the pattern of the
Wigner distribution. The squeezed distributions rotate along the
trajectories, which gives rise to the characteristic breathing
motion of its width. The phase space plots suggest that the two
involved wave packets roughly evolve on either of the potential
curves $V_-^D(x)$ or $V_+^A(x)$, and at the right side away from the
crossing point these two approximately coincide. We argued that the
corresponding classical times must be equal,
$T_{cl}^{A_+}=T_{cl}^{D_-}$, in order for the wave packets to
overlap at the curve crossing and interfere maximally. This,
however, does not imply that the longer time scales,
$T_{rev}^{A_�}$, $T_{rev}^{D_�}$, $T_{sup}^{A_�}, \ \ldots$,
of the potentials are equal, as will be discussed next.

\subsection{Long time evolution}\label{ssec3b}
We now turn to the dynamics over time scales of the order of
$T_{rev}$, which for our parameters correspond to several thousands
of classical oscillations. Since we are in the strong-excitation
regime, this does not, however, necessarily imply exceedingly long
total operational times.

\begin{figure}[ht]
\begin{center}
\includegraphics[width=8cm]{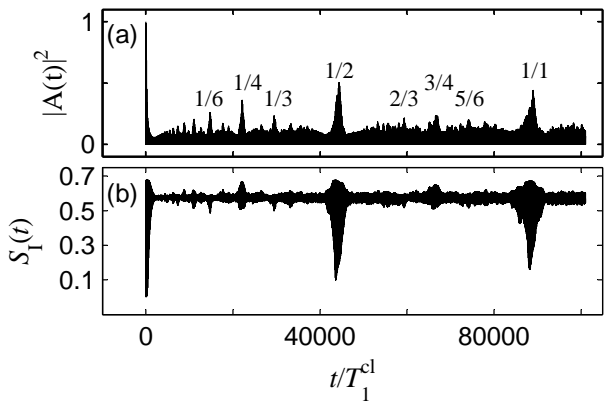}
\vspace{0cm} \caption{Autocorrelation function (a) and von Neumann
entropy (b) for an initial squeezed state in the case of 40-60
splitting.} \label{fig5}
\end{center}
\end{figure}

The expansion of the eigenvalues (\ref{eigexp}), in general,
contains an infinite number of terms, but for relatively smooth
potentials one expects fairly fast convergence. Higher order terms
typically cause imperfect full time and fractional revivals, by
which we mean that the amplitude of, say, the autocorrelation
function squared does not reach the values $1/j$, where $j=1,2,3,\
\ldots$, as expected \cite{wprev} for rational fractions of the
revival time $kT_{rev}/j$, $k$ and $j$ mutually prime. This effect,
arising from higher order anharmonicities, is naturally more
pronounced for longer times, and hence the revival amplitudes drop
for each consecutive revival period $T_{rev}$. The widths of the
revival peaks are getting broader, as well. As seen in
Fig.~\ref{fig3}, the uncertainty in phonon numbers $n$ affects the
collapse-revival pattern. In particular, the envelope functions of
the revival peaks are highly sensitive to the uncertainty $\Delta
n$, while the location of the peaks are determined by the average
$n_0$. In the ``ideal'' case, full amplitude revivals occur at
$t=jT_{rev}/2$ for integer $j$, half amplitude revivals at
$t=(2j+1)T_{rev}/4$, and so forth. In Fig.~\ref{fig5} we present the
autocorrelation function (a) and von Neumann entropy (b) for a time
span slightly longer than $T_{rev}$. The parameters and initial
conditions are similar to the ones of Figs.~\ref{fig3}.(d)-(f);
$\phi=1.07244080531656$ rad, $\lambda=0.064727653164347$,
$\sigma=0.0340999659$ and $\Delta=0.005025343787836$. The slight
change in parameters gives 40-60 splitting ($P_{sp}=0.4$) rather
than 50-50. Some of the numerous fractional revivals are labeled
conventionally \cite{wprev}. From Fig. \ref{fig5}.(b), we note that
in the collapse region the field and ion are highly entangled, but
the fact that $P_{sp}=0.4$ causes a decrease of entanglement
compared to the $P_{sp}=1/2$ case.

We have previously \cite{dong} studied the bistable motion and
long-term revivals in system Hamiltonians with more resemblance to
molecular models. In terms of the classical period, the revival time
found in the present study is much longer than those we calculated
for the molecular systems. This originates in a smaller
anharmonicity of the ion-trap system. In \cite{dong}, we found that
in accordance with expectations from a semiclassical derivation the
revival time of the coupled system $T_{rev}$ can be obtained from
the revival times of the adiabatic $T_{rev}^{A_+}$ and diabatic
$T_{rev}^{D_-}$ pathways and the wave packet splitting as
\begin{equation}\label{asarev}
\frac{1}{T_{rev}}=\frac{P_{sp}}{T_{rev}^{A_+}}+\frac{1-P_{sp}}{T_{rev}^{D_-}}.
\end{equation}
This relation was fulfilled for numerous systems, comprising both
coupled bound-bound and bound-repulsive states.
Equation~(\ref{asarev}) accurately reproduces the limiting cases of
adiabatic $P_{sp}=1$ and the diabatic $P_{sp}=0$ evolution. If
$T_{rev}^{A_+}\neq T_{rev}^{D_-}$, then we presumably have that the
size of $T_{rev}$ is between the values of $T_{rev}^{A_+}$ and
$T_{rev}^{D_-}$ in the intermediate range. However, in reality
$T_{rev}^{A_+}$, $T_{rev}^{D_-}$ and $P_{sp}$ are not independent
variables as they all depend on the system parameters in a complex
manner. To conclude, the revival time for multi-level systems can be
shorter than those of individual isolated systems. For the classical
periods, on the other hand, we found that
$T_{cl}=T_{cl}^{A_+}=T_{cl}^{D_-}$ for bistable trajectories. Thus,
the dynamics cannot be viewed as two uncoupled wave packets evolving
on the potential curves $V_+^A(x)$ and $V_-^D(x)$. Only interference
between these two wave packets can cause a common revival time
$T_{rev}$ shorter than one of $T_{rev}^{A_+}$ or $T_{rev}^{D_-}$.

\section{Conclusion}\label{sec4}
In this article we have studied in a non-standard way the dynamics
of a harmonically trapped ion pumped by a standing wave beyond the
RWA and the LD approximation. Using a wave packet technique, we
predict the existence of bistable states of motion similar to those
observed for molecular systems \cite{zhang,gado1,gado2,dong}. These
arise from interference of wave packets and hence it is a pure
quantum effect. Over longer
time periods these states possess a well resolved collapse-revival
pattern including full and fractional revivals.

Beyond the conjecture of the new class of bistable ion-trap states,
the article establishes a link between trapped ion systems and
molecular physics models. As such, we have put extra stress on how
the two areas relate to one another; expressing the model
Hamiltonian in $x$-representation and introducing diabatic and
adiabatic states and potential curves. We intend to further
investigate these directions, both by considering algebraic methods
in simple molecular models and by extending the current ion-trap
system to include additional ionic internal levels and vibrational
degrees of freedom. We will also allow for different trapping
potentials in order to more closely mimic molecular-like situations.
Anharmonic traps have been discussed in the literature, and then
mainly quadrupole trapping \cite{newtrap}, but in principal more
exotic shapes may be gained by reforming and adjusting the geometry
of the trap electrodes. On top, individual trapping potentials for
the
 internal ionic levels have been discussed \cite{hector2}.

\begin{acknowledgments}
This work was supported by EU-IP Programme SCALA (Contract No.
015714), the Swedish Research Council (VR) and NORDITA. Computer time at Uppsala
Multidiciplinary Center for Advanced Computational Sciences (UPPMAX)
is greatly acknowledged.
\end{acknowledgments}



\begin{thebibliography}{}
\bibitem{trapreview} D. J. Wineland, C. Monroe, W. M. Itano, D. Leibfried, B. E. King, and D. M. Meekhof, J.  Research Nat. Ins. Stand. Tech. {\bf 103}, 259 (1998); D. Leibfried, R. Blatt, C. Monroe, and D. Wineland, Rev. Mod. Phys. {\bf 75}, 281 (2003);W. P. Schleich, {\it Quantum Optics in Phase Space}, (Wiley, Berlin 2001); {\it Quantum Entanglement and Information Processing}, D. Esteve, J.-M. raimond, and J. Dalibard (Eds), (Elevier Amsterdam, 2004); S. Haroche, and J. M. Raimond, {\it Exploring the quantum}, (Oxford University Press, 2006).

\bibitem{expcat1} J. I. Cirac, A. S. Parkins, R. Blatt, and P. Zoller, Adv. Atom. Mol. Opt. Phys. {\bf 37}, 237 (1996).

\bibitem{traprev} D. M. Meekhof, C. Monroe, B. E. King, W. M. Itano, and D. J. Wineland, Phys. Rev. Lett. {\bf 76}, 1796 (1996).

\bibitem{trapent} S. Gulde, M. Riebe, G. P. T. Lancaster, C. Becher, J. Eschner, H. Haffner, F. Schmidt-Kaler, I. L. Chuang, and R. Blatt, Nature {\bf 421}, 48 (2003); D. Leibfried, B. DeMarco, V. Meyer, D. Lucas, M. Barrett, J. Britton, W. M. Itano, B. Jelenkovic, C. Langer, T. Rosenband, and D. J. Wineland, Nature {\bf 422}, 412 (2003); M. Riebe, H. Haffner, C. F. Roos, W. Hansel, J. Benhelm, G. P. T. Lancaster, T. W. Korber, C. Becher, F. Schmidt-Kaler, D. F. V. James, and R. Blatt, Nature {\bf 429}, 734 (2004); M. D. Barrett, J. Chiaverini, T. Schaetz, J. Britton, W. M. Itano, J. D. Jost, E. Knill, C. Langer, D. Leibfried, R. Ozeri, and D. J. Wineland, Nature {\bf 429}, 739 (2004).

\bibitem{expcat2} M. J. McDonnell, J. P. Home, D. M. Lucas, G. Imreh, B. C. Keitch, D. J. Szwer, N. R. Thomas, S. C. Webster, D. N. Stacey, and A. M. Steane, Phys. Rev. Lett. {\bf 98}, 063603 (2007).

\bibitem{hector1} H. Moya-Cessa, A. Vidiella-Barranco, J. A. Roversi, D. S. Freitas, and S. M. Dutra, Phys. Rev. A {\bf 59}, 2518 (1999); H. Moya-Cessa, D. Jonathan, and P. L. Knight, J. Mod. Opt. {\bf 50}265 (2003).

\bibitem{hector2} \"O. E. M\"ustecaplaglu, and L. You, Phys. Rev. A {\bf 65}, 033412 (2002).

\bibitem{expcav} A. B. Mundt, A. Kreuter, C. Becher *, D. Leibfried †, J. Eschner, F. Schmidt-Kaler, and R. Blatt, Phys. Rev. Lett. {\bf 89}, 103001 (2002)

\bibitem{theocav} J. I. Cirac, A. S. Parkins, R. Blatt, and P. Zoller, Opt. Comm. {\bf 97}, 353 (1993); V. Buzek, G. Drobn, M. S. Kim, G. Adam, and P. L. Knight, Phys. Rev. A {\bf 56}, 2352 (1997); X. B. Zou, K. Pahlke, and W. Mathis, Phys. Rev. A {\bf 65}, 064303 (2002).

\bibitem{bec-cav} P. Treutlein, D. Hunger, S. Camerer, T. W. H\"ansch, and J. Reichel, Phys. Rev. Lett. {\bf 99}, 140403 (2007); Y. Colombe, T. Steinmetz, G. Dubois, F. Linke, D. Hunger, and J. Reichel, Nature {\bf 450}, 272 (2007); F. Brennecke, T. Donner, S. Ritter, T. Bourdel, M. K\"ohn, and T. Esslinger, Nature {\bf 450}, 268 (2007); S. Slama, G. Krentz, S. Bux, C. Zimmermann, P. W. Courteille, Phys. Rev. A {\bf 75}, 063620 (2007).

\bibitem{ldprep} B. M. Rodriguez-Lara, H. Moya-Cessa, and A. B. Klimov, Phys. Rev. A {\bf 71}, 023811 (2005).

\bibitem{ldprep2} J. I. Cirac, A. S. Parkins, R. Blatt, and P. Zoller, Phys. Rev. Lett. {\bf 70}, 556 (1992); C. C. Gerry, Phys. Rev. A {\bf 55}, 2478 (1996).

\bibitem{ldprep3} L. X. Li, and G. -C. Guo, J. Opt. B: Quantum Semiclass. Opt. {\bf 1}, 339 (1999).

\bibitem{ldrev} J. I. Cirac, R. Blatt, A. S. Parkins, and P. Zoller, Phys. Rev. A {\bf 49}, 1202 (1993); M. F. Fang, S. Swain, and P. Zhou, Phys. Rev. A {\bf 63}, 013812 (2000); F. Mao-Fa, Chin. Phys. {\bf 11}, 1028 (2002).

\bibitem{ldrev2} C. A. Blockley, D. F. Walls, and H. Risken, Europhyss. Lett. {\bf 17}, 509 (1992).

\bibitem{ldmeas} S. Wallentowitz, and W. Vogel, Phys. Rev. Lett. {\bf 75}, 2932 (1995); C. D'Helon, and G. J. Milburn, Phys. Rev. A {\bf 54}, R25 (1995); F. E. Harrison, A. S. Parkins, M. J. Collett, and D. F. Walls, Phys. Rev. A {\bf 55}, 4412 (1997).

\bibitem{rwasw} Y. Wu, and X. Yang, Phys. Rev. Lett. {\bf 78}, 3086 (1997).

\bibitem{rwa2} S. Wallentowitz, W. Vogel and P.L. Knight, Phys.
Rev. A {\bf 59}, 531 (1999); S. Wallentowitz adn W. Vogel, Phys.
Rev. A {\bf 58}, 679 (1998).

\bibitem{rwaprep1} Q. Y. Yang, L. F. Wei, and L. E. Ding, J. Opt. B: Quantum Semiclass. Opt. {\bf 7}, 5 (2005).

\bibitem{rwaprep2} R. Blatt, J. I. Cirac, A. S. Parkins, and P. Zoller, Physica Scripta {\bf T59}, 294 (1995); R. L. de Matos, and W. Vogel, Phys. Rev. lett. {\bf 76}, 608 (1995).

\bibitem{rwaprep3} L.F. Wei, Y. Liu, and F. Nori, Phys. Rev. A {'bf 70}, 063801 (2004).

\bibitem{low} M. Feng. X. Zhu, X. Fang, M. Yan, and L. Shi, . Phys. B: At. Mol. Opt. Phys. {\bf 32}, 701 (1999).

\bibitem{lowprep} J. I. Cirac, R. Blatt, A. S. Parkins, And P. Zoller, Phys. Rev. Lett. {\bf 70}, 762 (1992).

\bibitem{strong} J. F. Poyatos, J. I. Cirac, R. Blatt, and P. Zoller, Phys. Rev. A, {\bf 54}, 1532 (1995); S. B. Zheng, X. W. Zhu, and M. Feng, Phys. Rev. A {\bf 62}, 033807 (2000).

\bibitem{ground} T. Liu, K. L. Wang, and M. Feng, J. Opt. B: At. Mol. Opt. Phys. {\bf 40}, 1967 (2007).

\bibitem{swprep2} S. B. Zheng, Phys. Lett. A {\bf 245}, 11 (1998).

\bibitem{qi} J. I. Cirac, and P. Zoller, Phys. Rev. Lett. {\bf 74}, 4091 (1995); K. M\o lmer, and A. S\o rensen, Phys. Rev. Lett. {\bf 82}, 1835 (1999); D. Jonathan, M. B. PLenio, and P. L. KNight, Phys. Rev. A {\bf 62}, 042307 (2000); J. I. Cirac, and P. Zoller, nature {\bf 404}, 579 (2000); F. Mintert, and C. Wunderlich, Phys. Rev. Lett. {\bf 87}, 257904 (2001).

\bibitem{oldwine} J. Larson, Phys. Scr. {\bf 76}, 146 (2007).

\bibitem{jonaswp} J. Larson, J. Phys.: Conf. Ser. {\bf 99} 012011 (2008).

\bibitem{heller75} E. J. Heller, \textit{J. Chem. Phys.}, {\bf 62}, 1544 (1975).

\bibitem{wpreview} B. M. Garraway, and K. A. Suominen,  Rep. Prog. Phys. {\bf 58} , 365 (1995).

\bibitem{dietz96} H. Dietz and V. Engel, \textit{Chem. Phys. Lett.}, {\bf 255}, 258 (1996).

\bibitem{zhang} B. Zhang, N. Gador, and T. Hansson, Phys. Rev. Lett. {\bf 91}, 173006 (2003).

\bibitem{gado1} N. Gador, B. Zhang, H. O. Karlsson, and T. Hansson, Phys. Rev. A {\bf 70}, 033418 (2004).

\bibitem{gado2} N. Gador, B. Zhang, and T. Hansson, Chem. Phys. Lett. {\bf 412}, 386 (2005).

\bibitem{dong} D. Wang, \AA. Larson, H. O. Karlsson, and T. Hansson, Chem. Phys. Lett. {\bf 449}, 266 (2007).

\bibitem{jc} E. T. Jaynes, and F. W. Cummings, Proc. IEEE {\bf 51}, 89 (1963); B. W. Shore, and P. L. Knight, J. Mod. Opt. {\bf 40}, 1195 (1993).

\bibitem{com2} Normally, resonance conditions on the involved frequencies determines the particular pumping (carrier or sidband) through the RWA.

\bibitem{com1} These dressed states differ from the ones defined in Subsection \ref{ssec2a}.

\bibitem{mankov} V. Mankov, G. Marmo, A. Porzio, S. Salimeno, and F. Zaccaria, Phys. Rev. A {\bf 62}, 053407 (2000).

\bibitem{noLD} W. Vogel, and R. L. de Matos Filho, Phys. Rev. A {\bf 52}, 4214 (1995).

\bibitem{wprev} R. W. Robinett, Phys. Rep. {\bf 392}, 1 (2004).

\bibitem{susy}  C. S. Jia, J. Y. Wang, S. He, and L. T. A. Sun, J. Phys. A: Math. Gen. {\bf 33}, 6993 (2000); N. F. Alexio, A. B. Balantekin, and M. A. C. Ribeiro, J. Phys. A: Math. Gen. {\bf 38}, 3173 (2000); S. Wallentowitz, I. A. Walmsley, L. J. Waxer, and Th. Richter, J. Phys. B: At. Mol. Opt. {\bf 35}, 1967 (2002); A. N. F. Alexio, and A. B. Balantekin, J. Phys. A: Math. Gen. {\bf 38}, 8603 (2005);  J. J. Pena, M. A. Romero-Romo, J. Morales, and J. L. Lopez-Bonilla, Int. J. Quant. Chem. {\bf 105}, 731 (2005).

\bibitem{lz} L. D. Landau, Z. Sowjet Union {\bf 2}, 46 (1932); C. Zener, \textit{Proc. R. Soc. (London) ser. A}, {\bf 137}, 696 (1932).

\bibitem{jonas1} I. Cusumano, A. Vaglicia, and G. Vetri, Phys. Rev. A {\bf 66}, 043408 (2002); J. Larson, J. Salo, and S. Stenholm, Phys. Rev. A {\bf 72}, 013814 (2005).


\bibitem{jcsup} P. F. G\'ora and C. Jedrzejek, Phys. Rev. A {\bf 48}, 3291 (1993); {\it idem}, Phys. Rev. A {\bf 49}, 3046 (1994).

\bibitem{cheby} H. Tal-Ezer and R. Kosloff, J. Chem. Phys. {\bf 81}, 3967 (1984).

\bibitem{araki} H. Araki, and E. Lieb, Comm. Math. Phys. {\bf 18}, 160 (1970).

\bibitem{jcrev} J. H. Eberly, N. B. Narozhny, and J.J. Sanchez-Mondragon, Phys. Rev. Lett. {\bf 44}, 1323 (1980); G. Rempe, H. Walther, and N. Klein, Phys. Rev. Lett. {\bf 58}, 353 (1987); M. Venkata Satyanarayana, P. Rice, R. Vyas, and H. J. Carmichael, J. Opt. Soc. Am. B {\bf 6}, 228 (1989); I. Sh. Averbukh, Phys. Rev. A {\bf 46}, R2205 (1992); M. Fleischauer, and W. P. Schleich, Phys. Rev. A {\bf 47}, 4258 (1993).

\bibitem{sq} These states are not nescecarily squeezed in its proper definition, but thier phase space distributions are rather elliptic than circular.

\bibitem{vitanov1} N. V. Vitanov, Phys. Rev. A {\bf 59}, 988 (1999).

\bibitem{jonas3} B. M. Garraway, and S. Stenholm, Phys. Rev. A {\bf 45}, 364 (1991); B. M. Garraway, and N. V. Vitanov, Phys. Rev. A {\bf 55}, 4418 (1996); J. Larson, Phys. Rev. A {\bf 73}, 013823 (2006).

\bibitem{wig} The way we define the Wigner distribution for multi-level systems by tracing out the ionic internal degrees of freedom (\ref{wigner}) naturally kills any information contained in the coherent crossing terms.

\bibitem{bumprev} E. Romera, and F. de los Santos, Phys. Rev. Lett. {\bf 99}, 263601 (2007).

\bibitem{newtrap} R. G. Brewer, R. G. DeVoe, and R. Kallenbach, Phys. Rev. A {\bf 46}, R6781 (1992); J. Walz, I. Siemers, M. Schubert, W. Neuhauser, R. Blatt, Europhys. Lett. {\bf 21}, 183 (1993).

\end{thebibliography}
\end{document}